\documentclass{amsart}
\usepackage{amsfonts}
\usepackage{amssymb}
\usepackage{graphicx}
\usepackage{amscd}

\newtheorem{theorem}{Theorem}
\theoremstyle{plain}

\numberwithin{equation}{section}

\vspace{5pt}

\begin{document}

\title[Bandwidth reduction in cognitive radio ]{Bandwidth reduction in
cognitive radio} \author{Tariq Shah$^{1}$, S. A. Hussain$^{2}$ and
Antonio Aparecido de Andrade$^{3}$} \thanks{$^{1}$Department of
Mathematics, Quaid-i-Azam University, Islamabad, Pakistan,
stariqshah@gmail.com, $^{2}$School of Engineering and Applied
Sciences (SEAS), ISRA University, Islamabad Campus, Pakistan,
1009-pd-ee002@isra.edu.pk and $^{3}$Department of Mathematics,
S\~{a}o Paulo State University at S\~{a}o Jos\'{e} do Rio Preto -
SP, Brazil, andrade@ibilce.unesp.br. Acknowledgment to FAPESP by
financial support, 2007/56052-8 and 2011/03441-2.}

\date{} \maketitle

\begin{abstract} Due to mushroom development of wireless devices cognitive radio is used to
resolve the bandwidth utilization and sacristy problem. The crafty
usage of bandwidth in cognitive radio based on error correcting
codes is ensured to accomodate un authorized user. This study
proposes a transmission model by which a finite sequence of binary
cyclic codes constructed by a binary BCH code of length $n=2^{s}-1$,
in which all codes have same error correction capability and code
rate but sequentially increasing code lengths greater than $n$.
Initially all these codes are carrying data of their corresponding
primary users. A transmission pattern is planned in the sprit of
interweave model deals the transmission parameters; modulation
scheme, bandwidth and code rate. Whenever, any of the primary users
having mod of transmission, the binary cyclic code, is not using its
allocated bandwidth, the user having its data built by binary BCH
code enter and exploit the free path as a secondary user. Eventually
whenever the primary user with $W$ bandwidth having binary BCH code
for its data transmission, change its status as a secondary user, it
just requires the bandwidth less than $W$. \end{abstract}

\subjclass{{\footnotesize \textbf{Mathematical subject
classification:} 11T71, 68P30, 94A15.}}

\keywords{{\footnotesize \textbf{Key words:} Bandwidth, secondary
user, primary user, BCH code, cyclic code, encoding, decoding,
Interweave.}}

\section{Introduction}

Cognitive radio is a most recent technology in wireless
communication by which the spectrum is vigorously utilized when the
primary user, the authorized holder of the spectrum, is not in. The
idea of cognitive radio is introduced in \cite{Mi}. Interpreting to
this notion the cognitive radio has the capability to evaluate the
radio surroundings and boost the decision permitting to the
transmission parameters such as modulation scheme, power, carrier
frequency, bandwidth and code rate. By \cite{BDG} power is allocated
to entire bandwidth for upsurge in capacity and keep the
interference at the primary user at the given start and endure the
complete transmission power inside properties.

Alternative inkling of the interference temperature model of \cite{CC} is
necessary for the primary receiver to dose the interference boundary,
henceforth the secondary user can transmit beneath the set level. The
fundamental plan in \cite{Z} is to matter license spectrum to secondary
users and guaranteed the interference perceived through primary users. To
protect the primary user commencing the interference triggered by the
secondary user through transmission, Srinivasa and Jafar \cite{SJ} offered
an organization of transmission models as: Interweave, overlay and underlay.

By \cite{Mi}, in the interweave model the secondary user utilized
the primary under utilized spectrum opportunistically and draw out
when the primary wants to in again. Thus the sensing is necessary
for the interweave scheme. However, dissimilar type of the sensing
procedures are used to identify the primary user and elude the
interference shaped by the secondary user. By \cite{K} for primary
user detection, three prominent methods known as: energy detection,
feature detection and match filter are getting attention. In
cognitive radio secondary user, permitted to use a spectrum of
primary user devoid of making interference to the primary user.
However, the secondary users need to recurrently monitor the
management of the spectrum to repel from interfering with the
primary user(see \cite{M}). Overlay and underlay stay with
spectrally modulated and spectrally encoding (SMSE) procedure all
beside with code division multiple access (CDMA) and orthogonal
frequency division multiple access (OFDMA). According to
\cite{SJCS}, in underlay, primary and secondary user transmit the
data concurrently under the situation that the secondary user
interferes less than an optimistic beginning with the primary user.
Spectrum sensing is not mandatory for underlay. Only the
interference limit is compulsory for the secondary user for talented
utilization of the spectrum. Henceforth, this interference
limitation confines the communication of secondary user to small
range. Whereas by \cite{BDG}, some range of the power should be
endorsed to the subcarrier of underlay as they also craft extra
interference to the primary user. The request of wireless devices is
increasing every day. Therefore, efficient utilization of spectrum
is a burring subject to decrease the spectrum over crowdedness. By
\cite{AA}, the overlay model permits the concurrent transmission of
primary and secondary users. The secondary transmitter is supposed
to have awareness of the primary message and use for sinking the
interference at its receiver. Precisely, in underlay and overlay
models the secondary users can transmit their data at once with the
primary users under some limitations: In underlay, secondary users
can transmit with enough power due to interference bound fixed at
the primary receiver whereas in the case of overlay transmission of
secondary users are solitary promising if secondary transmitter
recognizes the code-books and channel information. Besides, both of
these models do not guarantee that secondary user will not produce
interference for primary user during simultaneous transmission.
Similarly, the allowing surroundings for these models may damage the
transmission of both primary and secondary user.

In \cite{THXZB} wireless mesh network is practiced for terminal to
terminal bandwidth allocation which used the routing and scheduling
algorithms. The Max - min model is used for allocation of a fair
bandwidth. In the \cite{YBCMWR} TV band is utilized for the
cognitive radio over sensing and opportunistically use the vacant
frequencies. The diverse protocols are used for centralized and
decentralized spectrum distribution.

By using the undisclosed messages for cognitive radio channels first
of all fixed the boundaries for channel capacity. Two transmitters
having the primary and secondary messages go through the channel
which was received at the receivers with distinct primary and
secondary messages. By \cite{XJL}, improvement of spectral
efficiency in cognitive radio can be achieved by the secondary user
through the consent to utilize the untrustworthy part of the
spectrum which is assigned to the primary user. Henceforth the
optimal bandwidth is required from the numerous bandwidth allocated
to the primary and secondary users. Additionally, the secondary
users effort to choose the best possible bandwidth out of the
different many collections of bandwidths. Furthermore in \cite{XJL},
different spectrum sharing procedure is considered to improve the
cognitive radio networks.

In \cite{SMA} it is established that for a given binary BCH code
$C_{n}^{0}$ of length $n$ generated by a polynomial $g(x)$ of
$\mathbb{F}_{2}[x]$ of degree $r$ there exists a sequence of binary
cyclic codes $\{C_{2^{j-1}(n+1)n}^{j}\}_{j\geq 1}$ such that for
each $j\geq 1$, the binary cyclic code $C_{2^{j-1}(n+1)n}^{j}$ has
length $2^{j-1}(n+1)n$ and generated by $2^{j}r$ degree generalized
polynomial $g(x^{\frac{1}{2^{j}}})$ in
$\mathbb{F}_{2}[x,\frac{1}{2^{j}} \mathbb{Z}_{0}]$. Furthermore
$C_{n}^{0}$ is embedded in $C_{2^{j-1}(n+1)n}^{j}$ for each $j\geq
1$ and every code of the family $\{C_{2^{j-1}(n+1)n}^{j}\}_{j\geq
1}$ have same code rate and higher than of the binary BCH code
$C_{n}^{0}$. Furthermore, in \cite{SMA} a decoding algorithm is
proposed, by which the binary BCH code $C_{n}^{0}$ can be transmit
and decode through any of binary cyclic codes of the family
$\{C_{2^{j-1}(n+1)n}^{j}\}_{j\geq 1}$.

In the interweave model the secondary user exploited the primary
under utilized spectrum opportunistically and pull out when the
primary wants to in once again. Thus the sensing is necessary for
the interweave scheme. Though, dissimilar kind of the sensing
processes is used to detect the primary user and escape the
interference formed by the secondary user. For primary user
uncovering, bulging approaches, energy detection, feature detection
and match filter are receiving consideration.

Like interweave scheme, in this study we designed a transmission
model built on error correcting codes. Cognitive radio has the
ability to gauge the radio environs and boost the decision allowing
to the transmission parameters such as modulation scheme, bandwidth,
code rate, power, carrier frequency. In this study due to our model
formation we address the parameters; modulation scheme, bandwidth
and code rate. This model uses a finite embeddings sequence
$\{C_{2^{j-1}(n+1)n}^{j}\}_{j\geq 1}$ of binary cyclic codes
introduced in \cite{SMA} against a binary BCH code $C_{n}^{0}$ in
which all codes have same error correction capability and code rate
but sequentially increasing code lengths. Initially all these codes
including the binary BCH code are carrying data of their
corresponding primary users. The procedure advances as, whenever,
any of the primary users having mod of transmission, the binary
cyclic code, is not using its allocated bandwidth, the user having
its data configured by binary BCH code enter and utilize the free
path as a secondary user.

\section{Basic facts}

This section covers some of the basic results associated to monoid ring,
cyclic codes and specific on transmission parameters.

To construct a polynomial $(n,k)$-code $C$ over a finite Galois
field $\mathbb{F}_{q}$, where $q$ is power of some prime, we choose
a polynomial $g(x)$ of degree $n-k=r$ from $\mathbb{F}_{q}[x]$. A
message is represented by a polynomial, called the message
polynomial, $j(x)$ of degree less than or equal to $k-1$. The code
polynomial corresponding to this $j(x)$ is $v(x)$ and is equal to
$r(x)+x^{n-k}j(x)$, where $r(x)$ is the remainder of $x^{n-k}j(x)$
after dividing it by $g(x)$. A polynomial-code is an error
correcting code whose codewords consists of multiple of a given
fixed polynomial $g(x)$ known as the generator polynomial.

Let $(\mathbf{S},\ast )$ and $(\mathcal{R},+,.)$ be a commutative
semigroup and an associative unitary commutative ring respectively.
The set $\mathcal{SR}$ of all finitely nonzero functions $f$ from
$\mathbf{S}$ into $\mathcal{R}$ is a ring with respect to binary
operations addition and multiplication defined as
$(f+g)(s)=f(s)+g(s)$ and $(fg)(s)=\sum\limits_{t\ast u=s}f(t)g(u)$,
where the symbol $\sum\limits_{t\ast u=s}$ indicates that the sum is
taken over all pairs $(t,u)$ of elements of $\mathbf{S}$ such that
$t\ast u=s$ and it is settled that in the situation where $s$ is not
expressible in the form $t\ast u$ for any $t,u\in \mathbf{S}$,
$(fg)(s)=0$. The set $\mathcal{SR}$ is called a unitary commutative
semigroup ring of $\mathbf{S}$ over $\mathcal{R}$. If $\mathbf{S}$
is a monoid, then $\mathcal{SR}$ is called monoid ring. This ring
$\mathcal{SR}$ is represented as $\mathcal{R}[\mathbf{S}]$ whenever
$\mathbf{S}$ is a multiplicative semigroup and elements of
$\mathcal{T}$ are written either as $\sum\limits_{s\in
\mathbf{S}}f(s)s$ or as $\sum_{i=1}^{n}f(s_{i})s_{i}$. The
representation of $\mathcal{SR}$ will be $\mathcal{R}[x;\mathbf{S}]$
whenever $\mathbf{S}$ is an additive semigroup. A nonzero element
$f$ of $\mathcal{R}[x;\mathbf{S}]$ is uniquely represented in the
canonical form $\sum_{i=1}^{n}f(s_{i})x^{s_{i}}=
\sum_{i=1}^{n}f_{i}x^{s_{i}}$, where $f_{i}\neq 0$ and $s_{i}\neq
s_{j}$ for $i\neq j$. Of course, the monoid ring
$\mathcal{R}[x;\mathbf{S}]$ is a polynomial ring in one
indeterminate if $\mathbf{S}$ is the additive monoid
$\mathbb{Z}_{0}$ of non-negative integers.

The concept of degree and order are not generally defined in a
semigroup ring. Though if $\mathbf{S}$ is a totally ordered
semigroup, then the degree and order of an element of semigroup ring
$\mathcal{R}[x;\mathbf{S}]$ is defined as: if
$f=\sum_{i=1}^{n}f_{i}x^{s_{i}}$ is the canonical form of the
nonzero element $f\in \mathcal{R}[x;\mathbf{S}]$, where
$s_{1}<s_{2}<\cdots<s_{n}$, then $s_{n}$ is called the degree of $f$
and we write $\deg(f)=s_{n}$ and similarly the order of $f$ written
as $ord(f)=s_{1}$. Now, if $\mathcal{R}$ is an integral domain, then
for $f,g\in \mathcal{R}[x;\mathbf{S}]$, it follows that
$\deg(fg)=\deg(f)+\deg(g)$ and $ord(fg)=ord(f)+ord(g)$.

We start by an observation that, for a field $\mathbb{F}$ and an
integer $j\geq 0$, the structures of a polynomial ring
$\mathbb{F}[x]$ and a monoid ring
$\mathbb{F}[x,\frac{1}{j}\mathbb{Z}_{0}]$ have many
interconnections, for instance, for an ordered monoid $\mathbf{S}$,
the monoid ring $\mathbb{F}[x,\mathbf{S}]$ is a Euclidean domain if
$\mathbb{F}$ is a field and $\mathbf{S}\cong \mathbb{Z}$ or
$\mathbf{S}\cong \mathbb{Z}_{0}$ \cite[Theorem 8.4]{GP}. Of course
here $\frac{1}{j}\mathbb{Z}_{0}$ is totally ordered and and has an
isomorphism with $\mathbb{Z}_{0}$.

Let $\mathbb{F}$ be any field and $\frac{1}{j}\mathbb{Z}_{0}$ is the
additive monoid, then $\mathbb{F}[x;\frac{1}{j}\mathbb{Z}_{0}]$ is a
monoid ring. A generalized polynomial $g(x^{\frac{1}{j}})$ of
arbitrary degree $r$ in $\mathbb{F}[x;\frac{1}{j}\mathbb{Z}_{0}]$ is
represented as \begin{equation*}
g(x^{\frac{1}{j}})=g_{0}+g_{1,\frac{1}{j}}x^{\frac{1}{j}}+g_{2,\frac{1}{j}
}x^{\frac{1}{j}2}+\cdots+g_{r,\frac{1}{j}}x^{\frac{1}{j}r}
\end{equation*} Andrade and Shah has constructed cyclic codes over a local finite
commutative ring $\mathcal{R}$, through the monoid rings
$\mathcal{R}[x;\frac{1}{3}Z_{0}],\mathcal{R}[x;\frac{1}{2}Z_{0}]$
and $\mathcal{R}[x;\frac{1 }{2^{2}}Z_{0}]$ in \cite{ASK}, \cite{SKA}
and \cite{SKA1}, respectively. However, in \cite{MANAS} the cyclic
codes of certain types are discussed corresponding to the ascending
chain of monoid rings.

\section{Bandwidths of $n$ length BCH code and $2^{j-1}(n+1)n$ lengths
cyclic codes}

\subsection{Cyclic codes through $\mathbb{F}_{q}[x]$ and $\mathbb{F}_{q}[x;\frac{1}{2^{j}}
\mathbb{Z}_{0}]$}

For any positive integer $j$, there is a following ascending chain
of monoid rings given by \[ \mathbb{F}_{q}[x]\subset
\mathbb{F}_{q}[x;\frac{1}{2} \mathbb{Z}_{0}]\subset
\mathbb{F}_{q}[x;\frac{1}{2^{2}} \mathbb{Z}_{0}]\subset
\cdots\subset \mathbb{F}_{q}[x;\frac{1}{2^{j}}\mathbb{Z}_{0}]\subset
\cdots \]

Let $n=2^{s}-1$, where $s\in \mathbb{Z}^{+}$ and take
$(n+1)n=n^{\prime}$. Thus, $a_{0}+a_{\frac{1}{2^{j}}}\zeta
+a_{\frac{2}{2^{j}}}\zeta^{2}+\cdots+a_{\frac{(2^{j-1}n^{\prime
}-1)}{2^{j}}}\zeta^{2^{j-1}n^{\prime}-1}$ is a typical element of
$\mathbb{F}_{2}[x;\frac{1}{ 2^{j}}
\mathbb{Z}_{0}]/((x^{\frac{1}{2^{j}}})^{2^{j-1}n^{\prime}}-1)$,
where
$a_{0},a_{\frac{1}{2^{j}}},a_{\frac{2}{2^{j}}},\cdots,a_{\frac{_{(2^{j-1}n^{\prime
}-1)}}{2^{j}}}\in \mathbb{F}_{2}$ and $\zeta$ is the coset
$x^{\frac{1}{2^{j}}}+((x^{\frac{1}{2^{j}}})^{2^{j-1}n^{\prime
}}-1)$. So $f(\zeta)=0$, where $\zeta$ satisfies the relation
$\zeta^{2^{j-1}n^{\prime}}-1=0$.

Now, replace $x^{\frac{1}{2^{j}}}$ at the place of $\zeta$. Thus,
the ring $\mathbb{F}_{2}[x;\frac{1}{2^{j}}
\mathbb{Z}_{0}]/((x^{\frac{1}{2^{j}}})^{2^{j-1}n^{\prime}}-1)$
becomes $\mathbb{F}_{2}[x;\frac{1}{2^{j}} \mathbb{Z}_{\geq
0}]_{^{2^{j-1}n^{\prime}}}$ in which the relation
$(x^{\frac{1}{2^{j}}})^{2^{j-1}n^{\prime}}=1$ holds. The binary
operation multiplication $\ast$ in the ring
$\mathbb{F}_{2}[x;\frac{1}{2^{j}} \mathbb{Z}_{0}]_{{2^{j-1}n^{\prime
}}}$ is modulo the ideal
$((x^{\frac{1}{2^{j}}})^{2^{j-1}n^{\prime}}-1)$. So, given
$a(x^{\frac{1}{2^{j}} }),b(x^{\frac{1}{2^{j}}})\in
\mathbb{F}_{2}[x;\frac{1}{2^{j}}
\mathbb{Z}_{0}]_{{2^{j-1}n^{\prime}}}$, we write
$a(x^{\frac{1}{2^{j}}})\ast b(x^{\frac{1}{2^{j}}})$ to denote their
product in the ring $\mathbb{F}_{2}[x;\frac{1}{2^{j}}
\mathbb{Z}_{0}]_{{2^{j-1}n^{\prime}}}$ and
$a(x^{\frac{1}{2^{j}}})b(x^{\frac{1}{2^{j}}})$ to denote their
product in the ring $\mathbb{F}_{2}[x; \frac{1}{2^{j}}
\mathbb{Z}_{\geq 0}]$. If $\deg(a(x^{\frac{1}{2^{j}}}))+\deg(
b(x^{\frac{1}{2^{j}}}))< 2^{j-1}n^{\prime}$, then
$a(x^{\frac{1}{2^{j}}})\ast b(x^{\frac{1}{
2^{j}}})=a(x^{\frac{1}{2^{j}}})b(x^{\frac{1}{2^{j}}})$. Otherwise,
$a(x^{\frac{1}{2^{j}}})\ast b(x^{\frac{1}{2^{j}}})$ is the remainder
left on dividing $a(x^{\frac{1}{2^{j}}})b(x^{\frac{1}{2^{j}}})$ by
$(x^{\frac{1}{2^{j}}})^{2^{j-1}n^{\prime}}-1$. In other words, if
$a(x^{\frac{1}{2^{j}}})\ast
b(x^{\frac{1}{2^{j}}})=r(x^{\frac{1}{2^{j}}})$, then $a(x^{
\frac{1}{2^{j}}})b(x^{\frac{1}{2^{j}}})=r(x^{\frac{1}{2^{j}}})+((x^{\frac{1}{
2^{j}}})^{2^{j-1}n^{\prime}}-1)q(x^{\frac{1}{2^{j}}})$ for some
generalized polynomial $q(x^{\frac{1}{2^{j}}})$. Practically, to get
$a(x^{\frac{1}{2^{j}}})\ast b(x^{\frac{1}{2^{j}}})$, we only compute
the ordinary product $a(x^{\frac{1}{2^{j}}})b(x^{\frac{1}{2^{j}}})$
and then put $(x^{\frac{1}{2^{j}}})^{{\large 2}^{j-1}n^{\prime}}=1$,
$(x^{\frac{1}{2^{j}}})^{2^{j-1}(n+1)n+\frac{1}{2^{j}}}=x^{\frac{1}{2^{j}}}$,
and so on. Now, consider $x^{\frac{1}{2^{j}}}\ast
a(x^{\frac{1}{2^{j}}})$, and it would be \[
a_{\frac{(2^{j-1}n^{\prime}-1)}{2^{j}}}+a_{0}x^{\frac{1}{2^{j}}
}+a_{\frac{1}{2^{j}}}(x^{\frac{1}{2^{j}}})^{2}+\cdots+a_{\frac{(2^{j-1}n^{\prime
}-2)}{2^{j}}}(x^{\frac{1}{2^{j}}})^{2^{j-1}n^{\prime}-1}. \]

Particularly, take the product $x^{\frac{1}{2^{j}}}\ast
a(x^{\frac{1}{2^{j}} })$ in $\mathbb{F}_{2}[x;\frac{1}{2^{j}}
\mathbb{Z}_{0}]_{2^{j-1}n^{\prime}}$ as: the $\mathbb{F}_{2}$-space
$\mathbb{F}_{2}[x;\frac{1}{2^{j}}
\mathbb{Z}_{0}]_{2^{j-1}n^{\prime}}$ is isomorphic to
$\mathbb{F}_{2}$-space $\mathbb{F}_{2}^{2^{j-1}n^{\prime}}$, indeed,
$(x^{\frac{1}{2^{j}}})^{2^{j-1}n^{\prime}}-1=y^{2^{j-1}(n+1)n}-1$,
where $x^{\frac{1}{2^{j}}}=y$. In fact, we deal the coefficients of
generalized polynomials
$a(x^{\frac{1}{2^{j}}})=a_{0}+a_{\frac{1}{2^{j}}}x^{
\frac{1}{2^{j}}}+\cdots+a_{\frac{(2^{j-1}n^{\prime}-1)}{2^{j}}}(x^{
\frac{1}{2^{j}}})^{2^{j-1}n^{\prime}-1}$ of $\mathbb{F}_{2}[x;
\frac{1}{2^{j}} \mathbb{Z}_{0}]$, so $c(x^{\frac{1}{2^{j}}})$ has
$2^{j-1}n^{\prime}$ terms and hence the coefficients in
$\mathbb{F}_{2}$. Corresponding to the polynomial
$a(x^{\frac{1}{2^{j}}})$ of $\mathbb{F}_{2}[x;\frac{1}{2^{j}}
\mathbb{Z}_{0}]_{2^{j-1}n^{\prime}}$, there is a
$2^{j-1}n^{\prime}$-tuppled vector
$(a_{0},a_{\frac{1}{2^{j}}},\cdots,a_{\frac{(2^{j-1}n^{\prime
}-1)}{^{2^{j}}}})$ in $\mathbb{F}_{2}^{2^{j-1}n^{\prime}}$. Thus,
there is an isomorphism between the vector spaces
$\mathbb{F}_{2}[x;\frac{1}{2^{j}} \mathbb{Z}_{0}]_{2^{j-1}(n+1)n}$
and $\mathbb{F}_{2}^{2^{j-1}n^{\prime}}$, defined by $a\longmapsto
a(x^{\frac{1}{2^{j}}})$.

We observed that, multiplication by $x^{\frac{1}{2^{j}}}$ in the
ring $\mathbb{F}_{2}[x;\frac{1}{2^{j}} \mathbb{Z}_{\geq
0}]_{2^{j-1}n^{\prime}}$ corresponds to cyclic shift $\sigma$ in
$\mathbb{F}_{2}^{2^{j-1}n^{\prime}}$, that is,
$x^{\frac{1}{2^{j}}}\ast
a(x^{\frac{1}{2^{j}}})=\sigma(a)(x^{\frac{1}{2^{j}}})$.

A subspace $\mathcal{C}$ of $\mathbb{F}_{2}$-space
$\mathbb{F}_{2}^{2^{j-1}n^{\prime}}$ is a linear code. As already
agreed, we recognize every vector $\mathbf{a}$ in
$\mathbb{F}^{2^{j-1}n^{\prime}}$ with the polynomial
$a(x^{\frac{1}{2^{j}}})$ in $\mathbb{F}_{2}[x;\frac{1}{2^{j}}
\mathbb{Z}_{0}]_{2^{j-1}n^{\prime}}$, so the ideal
$\mathcal{C}_{2^{j-1}n^{\prime}}^{j}$ in
$\mathbb{F}_{2}[x;\frac{1}{2^{j}}
\mathbb{Z}_{0}]_{2^{j-1}n^{\prime}}$ is acyclic code. The elements
of the codes $\mathcal{C}_{2^{j-1}n^{\prime}}^{j}$ are now referred
as codewords or code generalized polynomials.

Note that if $\mathcal{C}_{2^{j-1}n^{\prime}}^{j}=(p(x^{\frac{1}{
2^{j}}}))$ is the ideal generated by $p(x^{\frac{1}{2^{j}}})$, then
$p(x^{\frac{1}{2^{j}}})$ is the generator (generalized) polynomial
of $\mathcal{C}_{2^{j-1}n^{\prime}}^{j}$ if and only if
$p(x^{\frac{1}{2^{j}}})$ is monic and divides
$(x^{\frac{1}{2^{j}}})^{2^{j-1}n^{\prime}}-1$ in
$\mathbb{F}_{2}[x;\frac{1}{2^{j}} \mathbb{Z}_{0}]$.

In \cite{SMA}, a link is developed between a binary BCH code
$(n,n-r)$ and a sequence
$\{2^{j-1}n^{\prime},2^{j-1}n^{\prime}-2^{j}r\}_{j\geq 1}$ of binary
cyclic codes.

Following \cite{SMA}, if $C_{n}^{0}$ is a binary BCH code $(n,n-r)$
based on the positive integers $c$, $\delta_{1}$, $q=2$ and $n$ such
that $2\leq \delta_{1}\leq n$ and $n=2^{s}-1$, where $s\in
\mathbb{Z}^{+}$. Thus, the binary BCH code $C_{n}^{0}$ has $r$
degree generator polynomial $g(x)={lcm}
\{j_{i}(x):i=c,c+1,\cdots,c+\delta_{1}-2\}$, where $j_{i}(x)$ are
minimal polynomials of $\zeta^{i}$, for
$i=c,c+1,\cdots,c+\delta_{1}-2$. Whereas $\zeta$ is the primitive
$n^{th}$ root of unity in $\mathbb{F}_{2^{l}}$. Since $j_{i}(x)$
divides $x^{n}-1$ for each $i$, it follows that $g(x)$ divides
$x^{n}-1$. Thus, $\mathcal{C}_{n}^{0}$ is a principal ideal in the
ring $\mathbb{F}_{2}[x]_{n}$ which is generated by $g(x)$.

\begin{theorem} \cite[Theorem 2]{SMA} For a positive integer $s$, let $C_{n}^{0}$ be
a binary BCH code of length $n=2^{s}-1$ generated by a polynomial
$g(x)=g_{0}+g_{1}x+\cdots+g_{r}x^{r}\in \mathbb{F}_{2}[x]$ of degree
$r$. Then, \begin{enumerate} \item there exists a sequence
$\{\mathcal{C}_{2^{j-1}n^{\prime}}^{j}\}_{j\geq 1}$ of binary cyclic
codes such that for each $j\geq 1$, the code
$\mathcal{C}_{2^{j-1}n^{\prime}}^{j}$ has length $2^{j-1}(n+1)n$,
generated by $2^{j}r$ degree generalized polynomial $g(x^{
\frac{1}{2^{j}}})=g_{0}+g_{1}(x^{\frac{1}{2^{j}}})^{2^{j}}+\cdots+g_{2^{j}r}(x^{
\frac{1}{2^{j}}})^{2^{j}r}\in \mathbb{F}_{2}[x,\frac{1}{2^{j}}
\mathbb{Z}_{0}]$, \item the binary BCH code $\mathcal{C}_{n}^{0}$ is
embedded in each binary cyclic code
$\mathcal{C}_{2^{j-1}n^{\prime}}^{j}$ for $j\geq 1$, \item there are
embeddings $\mathcal{C}_{n^{\prime}}^{1}\hookrightarrow
\mathcal{C}_{2^{1}n^{\prime}}^{2}\hookrightarrow \cdots
\hookrightarrow \mathcal{C}_{2^{j-1}n^{\prime}}^{j} \hookrightarrow
\cdots$ of binary cyclic codes of the sequence
$\{C_{2^{j-1}n^{\prime}}^{j}\}_{j\geq 1}$. \end{enumerate}
\end{theorem}

In \cite{SMA} it is established that for a binary BCH code
$\mathcal{C}_{n}^{0}=(g(x))$ there does not exist any binary BCH
code $\mathcal{C}_{2^{j-1}n^{\prime}}^{j}$ generated by polynomial
$g(x^{\frac{1}{2^{j}}})$.

The following table for
$\mathcal{C}_{2^{j-1}n^{\prime}}^{j}=(2^{j-1}n^{\prime},2^{j-1}n^{\prime}-
2^{j}r)$ can be constructed for varying, $n,r,j$, where integer
$j\geq 0$.

\begin{eqnarray*} &&\text{ \ \ \ \ \ \ \ \ \ \ \ \ \ \ \ \ \ \ \ \ \ \ \ \ \ \ \ \ \ \
\ \textbf{Table I \ \ \ \ \ }} \\ &&
\begin{tabular}{|l|l|l|l|l|l|l|} \hline $n,r$ & $C_{n}^{0}$ &
$C_{n^{\prime}}^{1}$ & $C_{2n^{\prime}}^{2}$ &
$C_{2^{2}n^{\prime}}^{3}$ & $C_{2^{3}n^{\prime}}^{4}$ & $\cdots$
\\ \hline $3,2$ & $(3,1)$ & $(12,8)$ & $(24,16)$ & $(48,32)$ &
$(96,64)$ & $\cdots$ \\ \hline \end{tabular} \end{eqnarray*}

\subsection{Bandwidth limitations}

Modulation is the process by which information is conveyed by means of an
electromagnetic wave. The information is impressed on a sinusoidal carrier
wave by varying its amplitude, frequency, or phase. Methods of modulation
may be either analog or digital. The power and bandwidth necessary for the
transmission of a signal with a given level of quality depends on the method
of modulation. There is a classic tradeoff between power and bandwidth that
is fundamental to the efficient design of communication systems. There are
three types of modulation; Amplitude shift keying (ASK), Frequency shift
keying (FSK), and Phase shift keying (PSK). Furthermore, quicker computer
processors allow the use of more complex forward error correction coding
techniques at high bit rates. Therefore, more spectrum proficient procedures
of digital modulation such as 8PSK and 16QAM are more gorgeous, even though
the power necessities are higher. Together with powerful coding methods such
as concatenated BCH coding, these methods offer the viewpoint of improved
spectral efficiency with essentially error-free digital signal transmission.

Let $S$ be the signal set $j$ be the number of signals in the signal
set. Suppose $v^{(t)}=(v_{0}^{(t)},\cdots,v_{n-1}^{(t)})\in
\mathbb{F}_{q}^{n}$ is the codeword of an $(n,k)$-code corresponding
to a message $u^{(t)}=(u_{0}^{(t)},\cdots,u_{k-1}^{(t)})\in
\mathbb{F}_{q}^{k}$ at time $t$ and we have divided each $v^{(t)}$
into $n/m$ blocks, where $m=log_{q}j$ and $j=q^{m}$ (the case
$q=2$). Thus, modulation is a map $M:\mathbb{F}_{q}^{m}\rightarrow
S$ defined as $s^{(t)}=s(v^{(t)})$, where $s^{(t)}\in S$ and $S$ is
a subset of $N$-dimensional real Euclidean space, that is, $S\subset
R^{N}$ \cite[Chapter 7]{KS}.

Following \cite{Rah}, the bandwidth required for an $(n,k)$ code is
$W=\frac{R_{u}}{m}(\frac{1}{R})$, where $m=\log _{2}M$, $R_{u}$ is
the source data (transmission) rate and $R=\frac{k}{n}$ the code
rate.

The bandwidth may be maximize and minimize, depends upon the minimum
and the maximum value of the $n/k$ $=1/R$ and the value of the, $m$
bits for selection of modulation scheme for different modulation
types. These bits may be minimum and maximum for maximum and minimum
bandwidth. It can be seen as
$W_{\max}=\frac{R_{u}}{m_{\min}}(\frac{1}{R})_{\max}$ and $W_{\min}=
\frac{R_{u}}{m_{\max}}(\frac{1}{R})_{\min}$. Thus, there are the
followings possibilities: \begin{enumerate} \item $m$ is fixed but
$\frac{1}{R}$ is varying, and \item $m$ and $\frac{1}{R}$ both are
varying. \end{enumerate}

For cognitive radio transformation under the interweave model we may
get spectrum corresponding to the given sequence
$\{\mathcal{C}_{2^{j-1}n^{\prime}}^{j}\}_{j=1}^{j_{0}}$ of binary
cyclic codes for data transfer of the primary users. Now, the setup
allow the secondary users having the binary BCH code
$\mathcal{C}_{n}^{0}$ mod for their data transfer. Accordingly the
secondary users obtain high speed data transfer as compare to its
own scheme of the BCH code $\mathcal{C}_{n}^{0}$. Furthermore, since
there are embeddings $\mathcal{C}_{n^{\prime}}^{1}\hookrightarrow
\mathcal{C}_{2^{1}n^{\prime}}^{2}\hookrightarrow
\cdots\hookrightarrow \mathcal{C}_{2^{j_{0}-1}n^{\prime}}^{j_{0}}$
of binary cyclic codes of the sequence
$\{\mathcal{C}_{{2^{j-1}n^{\prime}}^{j}}\}_{j=1}^{j_{0}}$ and the
binary BCH code $\mathcal{C}_{n}^{0}$ is embedded in each of binary
cyclic codes $\mathcal{C}_{2^{j-1}n^{\prime}}^{j}$ for $1\leq j\leq
j_{0}$.

It is noticed that corresponding to the code rate
$R_{n}^{0}=\frac{k}{n}$ of binary BCH code $\mathcal{C}_{n}^{0}$,
the code rate of binary cyclic code
$\mathcal{C}_{2^{j-1}n^{\prime}}^{j}$ is
$R_{2^{j-1}n^{\prime}}^{j}=\frac{2^{j-1}n^{\prime}-2^{j}r}{2^{j-1}n^{\prime}}$,
for each $1\leq j\leq j_{0}$. Moreover, $R_{n}^{0}\leq
R_{2^{j-1}n^{\prime}}^{j}$ and $R_{2^{j-1}n^{\prime}}^{j}$ is same
for each binary cyclic code $\mathcal{C}_{2^{j-1}n^{\prime}}^{j}$,
that is,
$R_{n^{\prime}}^{1}=R_{2n^{\prime}}^{2}=\cdots=R_{2^{j_{0}-1}n^{\prime}}^{j_{0}}$.
This means $1/R_{2^{j-1}n^{\prime}}^{j}\leq \frac{1}{R_{n}^{0}}$,
and therefore, $W_{2^{j-1}n^{\prime}}^{j}=\frac{R_{u}}{m}
1/R_{2^{j-1}n^{\prime}}^{j}\leq
W_{n}^{0}=\frac{R_{u}}{m}(1/R_{n}^{0})$. Thus, if we transmit data
through any of code of the sequence
$\{\mathcal{C}_{2^{j-1}n^{\prime}}^{j}\}_{j=1}^{j_{0}}$, the
bandwidth $W_{2^{j-1}n^{\prime}}^{j}$ for each $j\geq 1$ will be
lesser the bandwidth $W_{n}^{0}$ required for data transmitted
through the binary BCH code $\mathcal{C}_{n}^{0}$. Interestingly the
same bandwidth $W_{2^{j-1}n^{\prime}}^{j}=W$ is required for a user
having any type of code from the sequence
$\{\mathcal{C}_{2^{j-1}n^{\prime}}^{j}\}_{j=1}^{j_{0}}$ of binary
cyclic codes.

For $m=1,3$, the relation between bandwidth and code rate is given
as $W=w(R_{u}/m)(1/R)=wR_{u}/mR$, where $w$ is the bandwidth
expansion, $R_{u}$ is the transmission rate and $R=k/n$ is the code
rate. The bandwidth with different code rates is given in the Table
III (Chosen codes are from Table 1).

\begin{eqnarray*} &&\text{ \ \ \ \ \ \ \ \ \ \ \ \textbf{Table II}} \\
&& \begin{tabular}{|l|l|l|l|l|} \hline $R_{n}^{0}$ &
$R_{n^{\prime}}^{1}$ & $R_{2n^{\prime}}^{2}$ & $
R_{4n^{\prime}}^{3}$ & $R_{8n^{\prime}}^{4}$ \\ \hline $\frac{1}{3}$
& $\frac{2}{3}$ & $\frac{2}{3}$ & $\frac{2}{3}$ & $\frac{2}{3}$
\\ \hline \end{tabular} \end{eqnarray*}

For $w=1.2$ and $R_{u}=64$ $kbps$.\ \ \ \ \ \ \ \ \ \ \ \ \ \ \ \ \
\ \ \ \ \ \ \ \ \ \ \ \ \ \ \ \ \ \ \ \ \ \ \ \ \ \ \ \ \ \ \ \ \ \

\begin{eqnarray*}
&&\text{ \ \ \ \ \ \ \ \ \ \ \ \ \ \ \ \ \ \ \ \ \ \ \ \ \ \ \ \ \ \ \ \ \ \
\ \ \ \textbf{Table III}\ \ \ \textbf{\ }} \\
&& \begin{tabular}{|l|l|l|l|l|l|} \hline $m$ & $W_{n}^{0}\ kHz$ &
$W_{n^{\prime}}^{1}\ kHz$ & $W_{2n^{\prime}}^{2}\ kHz$ &
$W_{4n^{\prime}}^{3}\ kHz$ & $W_{8n^{\prime}}^{4} \ kHz$
\\ \hline $1$ & $W_{3}^{0}:236.4$ & $W_{12}^{1}:118.2$ &
$W_{24}^{2}:118.2$ & $ W_{48}^{3}:118.2$ & $W_{96}^{4}:118.2$ \\
\hline $3$ & $W_{3}^{0}:78.8$ & $W_{12}^{1}:39.4$ &
$W_{24}^{2}:39.4$ & $W_{48}^{3}:39.4$ & $W_{96}^{4}:39.4$ \\ \hline
\end{tabular} \end{eqnarray*}

\section{A transformation model for cognitive radio}

In communication Transmission Control Protocol (TCP) systematically
interlaced through RF front end, Physical layer, Data link layer,
Network layer, Transport Layer, and Upper Lyer. However, in this
study we deal Data link layer, which addresses the error correcting
codes.

Secondary user has an opportunistic accesses to the spectrum slum in the
interweave model whenever the primary user is not in and pull out when the
primary user desires to in once more. Therefore the codes constructed in
this study may provide an excellent scheme for wireless communication in
which interference issue is handled amicably. We propose a transmission
model for Cognitive radio based on error correcting codes which ensures the
non interference across the users.

Here in the following a sketch of the model is presented.

It is assumed that the primary users family
$\{\mathcal{P}_{2^{j-1}n^{\prime}}^{j}\}_{j=1}^{j_{0}}$
correspondingly use the family
$\{\mathcal{C}_{2^{{j-1}n^{\prime}}}^{j}\}_{j=1}^{j_{0}}$ of binary
cyclic codes for its data transmission. Since there are embeddings
\begin{equation*} \begin{array}{cccccc}
\mathcal{C}_{n^{\prime}}^{1} & \hookrightarrow &
\mathcal{C}_{2^{1}n^{\prime}}^{2} & \cdots & \hookrightarrow &
\mathcal{C}_{2^{j_{0}-1}n^{\prime}}^{j_{0}} \\
\circlearrowright &  & \circlearrowright &  &  & \circlearrowright \\
\mathcal{C}_{n}^{0} &  & \mathcal{C}_{n}^{0} &  &  &
\mathcal{C}_{n}^{0} \end{array} \end{equation*} of binary cyclic
codes of the sequence
$\{\mathcal{C}_{2^{j-1}n^{\prime}}^{j}\}_{j=1}^{j_{0}}$ and the
binary BCH code $\mathcal{C}_{n}^{0}$ is embedded in each of binary
cyclic code $\mathcal{C}_{2^{j-1}n^{\prime}}^{j}$ for $1\leq j\leq
j_{0}$.

Binary cyclic codes of the sequence
$\{\mathcal{C}_{2^{j-1}n^{\prime}}^{j}\}_{j=1}^{j_{0}}$ are used for
data transmission of the sequence
$\{\mathcal{P}_{2^{j-1}n^{\prime}}^{j}\}_{j=1}^{j_{0}}$ of primary
users with corresponding bandwidths
$\{W_{2^{j-1}n^{\prime}}^{j}\}_{j=1}^{j_{0}}$ such that
$W_{n^{\prime}}^{1}=W_{2n^{\prime}}^{1}=\cdots=W_{2^{j_{0}-1}n^{\prime}}^{j_{0}}=W$
and the total bandwidth $j_{{0}}W$  is required for simultaneous
transmission. As binary BCH code $\mathcal{C}_{n}^{0}$ requires
bandwidth $W_{n}^{0}\geq W$, so the data transmission of the
sequence $\{\mathcal{P}_{2^{j-1}n^{\prime}}^{j}\}_{j=1}^{j_{0}}$ of
primary users and the user $\mathcal{P}^{0}$ is $ j_{0}W+W_{n}^{0}$.

Whenever all users
$\{\mathcal{P}_{2^{j-1}n^{\prime}}^{j}\}_{j=1}^{j_{0}}$ and
$\mathcal{P}^{0}$ transmitting their data at a glance considered to
be the primary users. However, the user $\mathcal{P}^{0} $ enter as
a secondary user and opportunistically can use any of path of the
sequence $\{\mathcal{P}_{2^{j-1}n^{\prime}}^{j}\}_{j=1}^{j_{0}}$ of
primary users whenever any of them is not using its allotted
bandwidth. Here it is noticed that the data of the secondary user
$\mathcal{P}^{0}$ is configurated with the binary BCH code
$\mathcal{C}_{n}^{0}$ and it requires bandwidth higher than any of
the bandwidth required for the data of any primary user of the
sequence $\{\mathcal{P}_{2^{j-1}n^{\prime}}^{j}\}_{j=1}^{j_{0}}$.
Consequently, with high rate and with less bandwidth secondary user
$\mathcal{P}^{0}$ can transmit its data.

\subsection{How the model work} Notions

$j=0$, $1\leq j\leq 2^{j_{0}-1}n^{\prime}$.

$\mathcal{P}_{n}^{0}$: primary user corresponding to the binary BCH
code $\mathcal{C}_{n}^{0}$.

$\mathcal{P}_{2^{j-1}n^{\prime}}^{j}$: $j$-th primary user
corresponding to the binary cyclic code
$\mathcal{C}_{2^{j-1}n^{\prime}}^{j}$.

$m^{j}$: information symbols for j$th$ user.

$E^{j}$: $j$-th encoder for $m^{j}$.

$j_{\mathcal{P}^{j}}$: modulation for $E^{j}$.

$\blacksquare :$ bandwidth required for user
$\mathcal{P}_{2^{j-1}n^{\prime}}^{j}$ for each $j$.

$\blacklozenge \blacksquare$: bandwidth required for user
$\mathcal{P}_{n}^{0}$.

$\blacklozenge \blacksquare \blacksquare \blacksquare \blacksquare
\cdots\blacksquare$: $W_{n}^{0}+j_{0}W$, total bandwidth.

$Dj_{\mathcal{P}^{j}}$: $j$-th demodulation

$D^{j}$: $j$-th decoder.

The data of $\mathcal{P}^{j}$, for each $j$, is modulated through
\textbf{M}$_{\mathcal{P}^{j}}$, where \textbf{M}$_{\mathcal{P}^{j}}$
is a modulation map, i.e.,
\textbf{M}$_{\mathcal{P}^{j}}:\mathbb{F}_{q}^{m}\rightarrow
S_{\mathcal{P}^{j}}$, where $S_{\mathcal{P}^{j}}$ is the signal set,
$j_{\mathcal{P}^{j}}$ is the number of signals in the signal sets
$S_{\mathcal{P}^{j}}$. However, for $q=2$ it follows that
$j_{\mathcal{P}^{j}}=2^{m}$, where $m$ is a positive integer.

\begin{center} \underline{\textbf{A cognitive radio transmission
model}} \end{center}

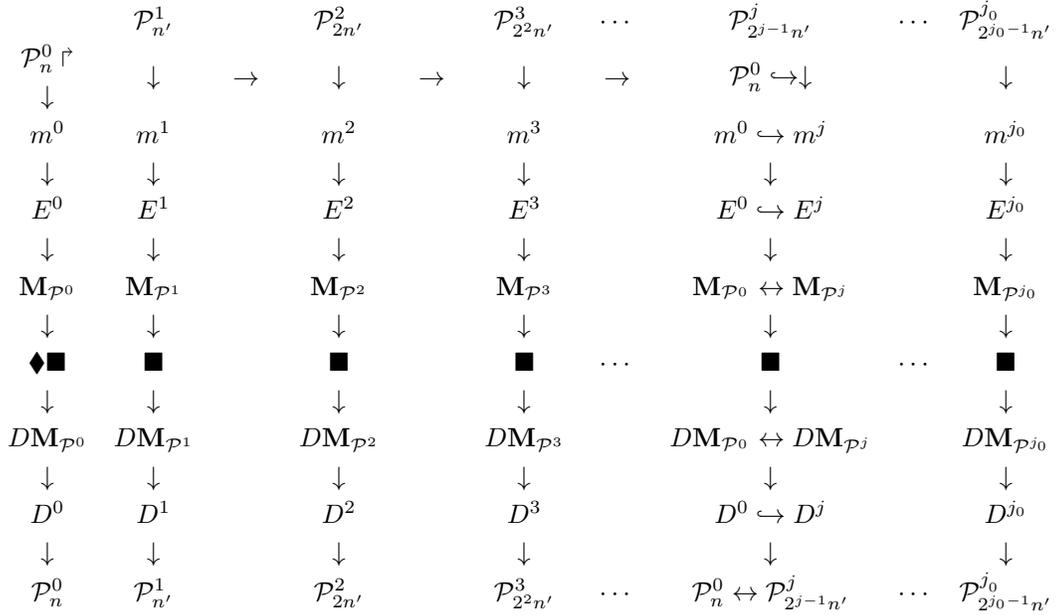
\begin{equation*} \begin{array}{cccccccccc}
& \mathcal{P}_{n^{\prime}}^{1} & & \mathcal{P}_{2n^{\prime}}^{2} & &
\mathcal{P}_{2^{2}n^{\prime}}^{3} & \cdots &
\mathcal{P}_{2^{j-1}n^{\prime}}^{j} & \cdots & \mathcal{P}_{2^{j_{0}-1}n^{\prime}}^{j_{_{0}}} \\
\begin{array}{c} \mathcal{P}_{n}^{0}\Rsh \\ \downarrow \end{array} & \downarrow &
\begin{array}{c} \rightarrow \\ \end{array} & \downarrow &
\begin{array}{c} \rightarrow \\ \end{array} & \downarrow &
\begin{array}{c} \rightarrow \\ \end{array} & \mathcal{P}_{{\large n}}^{0} \hookrightarrow \downarrow & & \downarrow \\
m^{0} & m^{1} & & m^{2} & & m^{3} & & m^{0} \hookrightarrow m^{j} &
& m^{j_{0}} \\ \downarrow & \downarrow & & \downarrow & & \downarrow
& & \downarrow & & \downarrow \\ E^{0} & E^{1} & &
E^{2} & & E^{3} & & E^{0} \hookrightarrow E^{j} & & E^{j_{0}} \\
\downarrow & \downarrow & & \downarrow & & \downarrow & & \downarrow
& & \downarrow \\ \mathbf{M}_{\mathcal{P}^{0}} &
\mathbf{M}_{\mathcal{P}^{1}} & & \mathbf{M}_{\mathcal{P}^{2}} & &
\mathbf{M}_{\mathcal{P}^{3}} & & \mathbf{M}_{\mathcal{P}^{_{0}}}
\leftrightarrow \mathbf{M}_{\mathcal{P}^{j}} & & \mathbf{M}_{\mathcal{P}^{j_{0}}} \\
\downarrow & \downarrow & & \downarrow & & \downarrow & & \downarrow
& & \downarrow \\ \blacklozenge \blacksquare & \blacksquare & &
\blacksquare & & \blacksquare & \cdots & \blacksquare & \cdots & \blacksquare \\
\downarrow & \downarrow & & \downarrow & & \downarrow & & \downarrow
& & \downarrow \\ D\mathbf{M}_{\mathcal{P}^{0}} &
D\mathbf{M}_{\mathcal{P}^{1}} & & D\mathbf{M }_{\mathcal{P}^{2}} & &
D\mathbf{M}_{\mathcal{P}^{3}} & & D\mathbf{M}_{
\mathcal{P}^{_{0}}}\leftrightarrow D\mathbf{M}_{\mathcal{P}^{j}} & &
D\mathbf{M}_{\mathcal{P}^{j_{0}}} \\ \downarrow & \downarrow & &
\downarrow & & \downarrow & & \downarrow & & \downarrow \\
D^{0} & D^{1} & & D^{2} & & D^{3} & & D^{0} \hookrightarrow D^{j} &
& D^{j_{0}} \\ \downarrow & \downarrow & & \downarrow & & \downarrow
& & \downarrow & & \downarrow \\ \mathcal{P}_{n}^{0} &
\mathcal{P}_{n^{\prime}}^{1} & & \mathcal{P}_{2n^{\prime}}^{2} & &
\mathcal{P}_{2^{2}n^{\prime}}^{3} & \cdots & \mathcal{P}_{n}^{0}
\leftrightarrow \mathcal{P}_{2^{j-1}n^{\prime}}^{j} & \cdots &
\mathcal{P}_{2^{j_{0}-1}n^{\prime}}^{j_{_{0}}} \end{array}
\end{equation*}

{\bf Transmission steps}: Let $j=0$ and $1\leq j\leq j_{0}$.
\begin{enumerate} \item \textbf{All users are primary users}:
\begin{enumerate} \item Data of the $\mathcal{P}^{j}$, for each $j$, users transform into
the set $m^{j}$ of message bits. \item For each $j$, the set $m^{j}$
of message bits encoded through encoder $E^{j}$. \item For each $j$,
the set $E^{j}$ of encoded messages modulated through
\textbf{M}$_{\mathcal{P}^{j}}$. \item For each $j$, the set
\textbf{M}$_{\mathcal{P}^{j}}$ of modulated codewords passing
through the channel having bandwidth $W^{j}=W$. \item For each $j$,
the corresponding transmitted signals of
\textbf{M}$_{\mathcal{P}^{j}}$ are demodulated. \item For each $j$,
the received signals corresponding to \textbf{M}$_{\mathcal{P}^{j}}$
are decoded through decoder $D^{j}$. \item The end of whole process
is the destination of data of all users. \end{enumerate} \item
\textbf{All users are not primary users}: almost all steps of data
transmission are same as I. Though, the user $ \mathcal{P}^{0}$
enter as a secondary user and opportunistically can use any of the
path of the sequence
$\{\mathcal{P}_{2^{j-1}n^{\prime}}^{j}\}_{j=1}^{j_{0}}$ of primary
users whenever any of them is not using its allotted bandwidth. For
instance, if the primary user $\mathcal{P}_{2^{j-1}n^{\prime}}^{j}$
is not in, then the user $\mathcal{P}^{0}$ transmitted its data
configurated by the binary BCH code $\mathcal{C}_{n}^{0}$ through
the binary cyclic code $\mathcal{C}_{2^{j-1}n^{\prime}}^{j}$ used
for data of primary user $\mathcal{P}_{2^{j-1}n^{\prime}}^{j}$.
\end{enumerate}

\subsection{Illustration}

Let $n=2^{2}-1=3$, $\delta =3$, $c=1$, $p(x)=x^{2}+x+1$ a primitive
polynomial of degree $2$ and
$\mathbb{F}_{2^{2}}=\frac{\mathbb{F}_{2}[x]}{(p(x))}=\{a_{0}+a_{1}\zeta
:a_{0},a_{1}\in \mathbb{F}_{2}\}$, where $\zeta$ satisfies the
relation $\zeta^{2}+\zeta+1=0$. Using this relation, we obtain
$\{0,\zeta,1+\zeta\}$. Let $m_{i}(x)$ be the minimal polynomial of
$\zeta ^{i}$, for $i=c,c+1,\cdots,c+\delta-2$. Thus,
$m_{1}(x)=x^{2}+x+1$ and hence
$g(x)={lcm}\{m_{i}(x):i=c,c+1,\cdots,c+ \delta-2\}=x^{2}+x+1$. The
code $C_{3}^{0}=(g(x))\subset \mathbb{F}_{2}[x]_{{3}}$ is a binary
BCH code based on the positive integers $c=1$, $\delta=3$, $q=2$ and
$n=3$ such that $2\leq \delta \leq n$ with $gcd(n,2)=1$.
Consequently, $\{\mathcal{C}_{2^{j-1}n^{\prime}}^{j}\}_{j\geq
1}=\{\mathcal{C}_{2^{j-1}(3+1)3}^{j}\}_{j\geq
1}=\{\mathcal{C}_{3\times 2^{j+1}}^{j}\}_{j\geq 1}$ is the sequence
$\{(3\times 2^{j+1},3\times 2^{j+1}-2^{j+1})\}_{j\geq 1}$ of binary
cyclic codes corresponding to the BCH code $\mathcal{C}_{3}^{0}$.
For instance
$\mathcal{C}_{2^{1-1}(3+1)3}^{1}=\mathcal{C}_{{12}}^{1}$, a $(12,8)$
code which is generated by
$g(x^{\frac{1}{2}})=(x^{\frac{1}{2}})^{2(2)}+(x^{
\frac{1}{2}})^{2}+1=(x^{\frac{1}{2}})^{4}+(x^{\frac{1}{2}})^{2}+1$
in $\mathbb{F}_{2}[x,\frac{1}{2^{1}} \mathbb{Z}_{0}]$ and
$C_{2^{2-1}(3+1)3}^{2}=C_{24}^{2}$ is a $(24,16)$ code which is
generated by $g(x^{\frac{1}{2^{2}}})=(x^{\frac{1}{
2^{2}}})^{2^{2}(2)}+(x^{\frac{1}{2^{2}}})^{2^{2}}+1=(x^{\frac{1}{2^{2}}
})^{8}+(x^{\frac{1}{2^{2}}})^{4}+1$ in
$\mathbb{F}_{2}[x,\frac{1}{2^{2}} \mathbb{Z}_{0}]$, and so on.

Follow the Table I and label the corresponding bandwidths as:

\begin{eqnarray*} \text{For} m &=&1 \\
W_{3}^{0} &:&236.4:\blacksquare \blacksquare \\
W_{12}^{1} &=&W_{24}^{2}:118.2:\blacksquare
\end{eqnarray*}

\begin{eqnarray*} \text{For} m &=&3 \\
W_{3}^{0} &:&69.78:\blacksquare \blacksquare \\
W_{12}^{1} &=&W_{24}^{2}:34.89:\blacksquare
\end{eqnarray*}

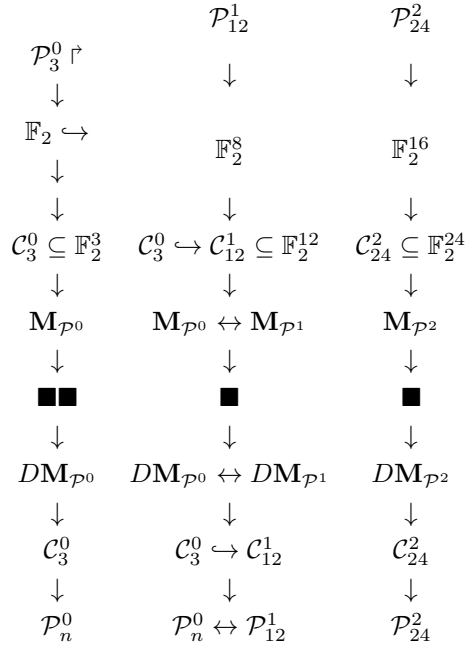
\begin{equation*} \begin{array}{ccc}
& \mathcal{P}_{12}^{1} & \mathcal{P}_{24}^{2} \\ \begin{array}{c}
\mathcal{P}_{3}^{0}\Rsh \\ \downarrow
\end{array} & \downarrow & \downarrow \\ \begin{array}{c}
\mathbb{F}_{2}\hookrightarrow \\ \downarrow \end{array} & \mathbb{F}_{2}^{8} & \mathbb{F}_{2}^{16} \\
\downarrow & \downarrow & \downarrow \\ \mathcal{C}_{3}^{0}\subseteq
\mathbb{F}_{2}^{3} & \mathcal{C}_{3}^{0}\hookrightarrow
\mathcal{C}_{12}^{1}\subseteq \mathbb{F}_{2}^{12} & \mathcal{C}_{24}^{2}\subseteq \mathbb{F}_{2}^{24} \\
\downarrow & \downarrow & \downarrow \\ \mathbf{M}_{\mathcal{P}^{0}}
& \mathbf{M}_{\mathcal{P}^{0}}\leftrightarrow \mathbf{M}_{\mathcal{P}^{1}} & \mathbf{M}_{\mathcal{P}^{2}} \\
\downarrow & \downarrow & \downarrow \\ \blacksquare \blacksquare & \blacksquare & \blacksquare \\
\downarrow & \downarrow & \downarrow \\
D\mathbf{M}_{\mathcal{P}^{0}} & D\mathbf{M}_{\mathcal{P}^{0}}
\leftrightarrow D\mathbf{M}_{\mathcal{P}^{1}} & D\mathbf{M}_{\mathcal{P}^{2}} \\
\downarrow & \downarrow & \downarrow \\ \mathcal{C}_{3}^{0} &
\mathcal{C}_{3}^{0}\hookrightarrow \mathcal{C}_{12}^{1} & \mathcal{C}_{24}^{2} \\
\downarrow & \downarrow & \downarrow \\ \mathcal{P}_{n}^{0} &
\mathcal{P}_{n}^{0} \leftrightarrow \mathcal{P}_{12}^{1} &
\mathcal{P}_{24}^{2} \end{array} \end{equation*}

If the data of user $\mathcal{P}_{3}^{0}$ is transmitted through the
binary BCH code $C_{3}^{0}$ for any modulation scheme, it requires
double bandwidth than the bandwidth required for its transmission
through the path of any of the binary cyclic codes $C_{12}^{0}$ and
$C_{24}^{0}$.

In the following we illustrate the decoding steps. Consider the
binary BCH code $C_{3}^{0}$.

The canonical generator matrix of binary cyclic $(12,8)$ code
$C_{12}^{1}$ is given by \begin{equation*} G^{^{1}}=\left[
\begin{array}{cccccccccccc} 1 & 0 & 1 & 0 & 1 & 0 & 0 & 0 & 0 & 0 & 0 & 0 \\
0 & 1 & 0 & 1 & 0 & 1 & 0 & 0 & 0 & 0 & 0 & 0 \\
0 & 0 & 1 & 0 & 1 & 0 & 1 & 0 & 0 & 0 & 0 & 0 \\
0 & 0 & 0 & 1 & 0 & 1 & 0 & 1 & 0 & 0 & 0 & 0 \\
0 & 0 & 0 & 0 & 1 & 0 & 1 & 0 & 1 & 0 & 0 & 0 \\
0 & 0 & 0 & 0 & 0 & 1 & 0 & 1 & 0 & 1 & 0 & 0 \\
0 & 0 & 0 & 0 & 0 & 0 & 1 & 0 & 1 & 0 & 1 & 0 \\
0 & 0 & 0 & 0 & 0 & 0 & 0 & 1 & 0 & 1 & 0 & 1
\end{array} \right], \end{equation*} which is obtained by the generator polynomial
$g(x^{\frac{1}{2}})=1+(x^{\frac{1}{2}})^{2}+(x^{\frac{1}{2}})^{4}$,
whereas the parity check-matrix with check polynomial
$h(x^{\frac{1}{2}})=1+(x^{\frac{1}{2}})^{2}+(x^{\frac{1}{2}
})^{6}+(x^{\frac{1}{2}})^{8}$ is given \begin{equation*}
H^{{1}}=\left[ \begin{array}{cccccccccccc} 1 & 0 & 1 & 0 & 0 & 0 & 1 & 0 & 1 & 0 & 0 & 0 \\
0 & 1 & 0 & 1 & 0 & 0 & 0 & 1 & 0 & 1 & 0 & 0 \\
0 & 0 & 1 & 0 & 1 & 0 & 0 & 0 & 1 & 0 & 1 & 0 \\
0 & 0 & 0 & 1 & 0 & 1 & 0 & 0 & 0 & 1 & 0 & 1
\end{array} \right]. \end{equation*} Syndrome table is given by
\begin{equation*} \begin{array}{ccc} & \text{\textbf{coset leader}} & \text{syndrome} \\
e_{0} & 000000000000 & 0000 \\ e_{1} & 100000000000 & 1000 \\
e_{2} & 010000000000 & 0100 \\ e_{3} & 001000000000 & 1010 \\
e_{4} & 000100000000 & 0101 \\ e_{5} & 000010000000 & 0010 \\
e_{6} & 000001000000 & 0001 \\ e_{7} & 110000000000 & 1100 \\
e_{8} & 100100000000 & 1101 \\ e_{9} & 100001000000 & 1001 \\
e_{10} & 011000000000 & 1110 \\ e_{11} & 010010000000 & 0110 \\
e_{12} & 001100000000 & 1111 \\ e_{13} & 001001000000 & 1011 \\
e_{14} & 000110000000 & 0111 \\ e_{15} & 000011000000 & 0011
\end{array} \end{equation*} Let $b=101\in \mathbb{F}_{2}^{3}$ be the received vector of binary BCH code $C_{3}^{0}$. Thus, its polynomial representation is
$b(x)=1+x^{2}$ in $\mathbb{F}_{2}[x]_{{3}}$ and corresponding
received polynomial in the cyclic code $C_{12}^{1}$ is given by
$b^{{{\prime}}}(x^{\frac{1}{2}})=1+(x^{\frac{1}{2}})^{4}$ in
$\mathbb{F}_{2}[x;\frac{1}{2} \mathbb{Z}_{0}]_{{12}}$ by using
\cite[Theorem 1]{SMA}, and its vector representation will be
$b^{{\prime}}=100010000000$ in $\mathbb{F}_{2}^{12}$ and
$(b^{{\prime}})=b^{{\prime}}(H^{{1}})^{T}=1010=S(e_{3})$. Hence, the
corrected codeword in $C_{12}^{1}$ is
$a^{{\prime}}=b^{{\prime}}+e_{3}=101010000000 $ and its polynomial
representation is
$a^{{\prime}}(x^{\frac{1}{2}})=1+(x^{\frac{1}{2}})^{2}+(x^{\frac{1}{2}})^{2(2)}$
in $\mathbb{F}_{2}[x;\frac{1}{2} \mathbb{Z}_{0}]_{12}$. Thus, the
corresponding corrected codeword in binary BCH code $C_{3}^{0}$ is
$a(x)=1+x+x^{2}$ in $\mathbb{F}_{2}[x]_{_{3}}$, i.e., $a=111$.

In the similar fashion we can decode received vector of $C_{3}^{0}$
through the decoding of any member of the family
$\{C_{2^{j-1}(n+1)n}^{j}\}_{j\geq 1}$ instead of $C_{12}^{1}$.

\section{Conclusion}

This study proposed a novel interweave inclined transmission model
for cognitive radio. The data of primary user $\mathcal{P}^{0}$ is
configurated and transmitted through the binary BCH code
$\mathcal{C}_{n}^{0}$. However, the data of the family
$\{\mathcal{P}_{2^{j-1}n^{\prime}}^{j}\}_{j=1}^{j_{0}}$ of primary
users is configurated by the family
$\{\mathcal{C}_{2^{j-1}n^{\prime}}^{j}\}_{j=1}^{j_{0}}$ having
sequentially increasing code lengths but with same code rate. Due to
the modulation scheme every member of
$\{\mathcal{C}_{2^{j-1}n^{\prime}}^{j}\}_{j=1}^{j_{0}}$ requires
same bandwidth but lesser than required for the binary BCH code
$\mathcal{C}_{n}^{0}$.

Initially all these codes are carrying data of their corresponding
primary users. A transmission pattern is planned in the spirit of
interweave model in such a way that the user $\mathcal{P}^{0}$
observe and opportunistically avail the channel path of any of the
Primary users of the family
$\{\mathcal{P}_{2^{j-1}n^{\prime}}^{j}\}_{j=1}^{j_{0}}$ not
utilizing its allotted bandwidth.

This study can also be extended for a set of different $n$, the
length of the binary BCH code $\mathcal{C}_{n}^{0}$. Consequently, a
multiple transformation model can be designed.

\end{document}